\documentclass[11pt]{article}

\usepackage{amsxtra,amssymb,amsthm,amsmath,latexsym}
\usepackage{graphicx}
\usepackage{epsfig}
\usepackage{afterpage}

\newcommand{\R}{{\mathbb R}}

\newcommand{\dl}{{\delta}}
\newcommand{\bee}{\begin{equation*}}
\newcommand{\eee}{\end{equation*}}
\newcommand{\be}{\begin{equation}}
\newcommand{\ee}{\end{equation}}

\title{Many-body wave scattering problems in the case of small scatterers
}
\author{Alexander G. Ramm \\
\small Department of Mathematics\\[-0.8ex]
\small Kansas State University, Manhattan, KS 66506-2602, USA\\
\small \texttt{ramm@math.ksu.edu}}

\date{}
\begin{document}

\maketitle
\begin{abstract}Formulas are derived for solutions of many-body wave
scattering problems by small particles in the case of acoustically
soft, hard, and impedance particles embedded in an inhomogeneous
medium. The case of transmission (interface) boundary conditions is
also studied in detail. The
limiting case is considered, when the size $a$ of small
particles tends to zero while their number tends to infinity at a
suitable rate. Equations for the limiting effective
(self-consistent) field in the medium are derived.
The theory is based on a study of integral equations and
asymptotics of their solutions as $a\to 0$. The case of
wave scattering by many small particles embedded in an inhomogeneous
medium is also studied.
\end{abstract}



\section{Introduction}
There is a large literature on wave scattering by small bodies,
starting from Rayleigh's work (1871), \cite{Ray}, \cite{LL}, \cite{H}. For
the problem of wave
scattering by one body an analytical solution was found only for the
bodies of special shapes, for example, for balls and ellipsoids. If
the scatterer is small then the scattered field can be calculated
analytically for bodies of arbitrary shapes, see \cite{R476}, where
this theory is presented.

The many-body wave scattering problem was discussed in the
literature mostly numerically, if the number of scatterers is small,
or under the assumption that the influence of the waves, scattered
by other particles on a particular particle is negligible (see
\cite{M}, where one finds a large bibliography, 1386 entries). This
corresponds to the case when the distance $d$ between neighboring
particles is much larger than the wavelength $\lambda$, and the
characteristic size $a$ of a small body (particle) is much smaller
than $\lambda$. By $k=\frac{2\pi}{\lambda}$ the wave number is
denoted.

This paper is a review of the author's results. The results
of  papers \cite{R607} and \cite{R615} are presented. The presentation
follows closely the above papers.

The basic results of this paper consist of:

i) Derivation of analytic formulas for the scattering amplitude for the
wave
scattering problem by one small ($ka\ll1$) body {\it of an arbitrary
shape} under the Dirichlet, impedance, and  Neumann boundary condition
(acoustically soft, impedance, and hard particle), and the
transmission (interface) boundary condition;

ii) Solution to {\it many-body wave
scattering problem} by small particles, embedded in an inhomogeneous
medium,  under the assumptions
$a\ll d$ and $a\ll \lambda$, where $d$ is the minimal distance
between neighboring particles;

iii) Derivation of the equations for
the limiting effective (self-consistent) field in an inhomogeneous
 medium in which many small particles are embedded, when
$a\to 0$ and the number $M=M(a)$ of the small particles tends to
infinity at an appropriate rate;

iv) Derivation of linear algebraic
systems for solving many-body wave scattering problems. These systems
are not obtained by a discretization of boundary integral equations,
and they give an efficient numerical method for solving many-body wave
scattering problems in the case of small scatterers;

v) Theory of wave scattering by small bodies of an arbitrary
shape under the transmission (interface) boundary condition,
and a derivation of the equation for the effective field in the
limiting medium consisting of very many small bodies.

The derivations of the results, presented in this paper, are
rigorous. The order of the error estimates as $a\to 0$ is obtained.
Our methods give powerful numerical methods for solving many-body
wave scattering problems in the case when the scatterers are small
(see \cite{R609}, \cite{R625}).

In Sections 1-4 wave scattering by small bodies under the Dirichlet,
Neumann, and impedance boundary conditions is developed.
In Sections 5-7 wave scattering by small bodies under the transmission
boundary condition is presented. In Section 8 Conclusions are briefly
stated.

Let us formulate the wave scattering problems we deal with. First,
let us consider a one-body scattering problem. Let $D_1$ be a bounded
domain in $\R^3$ with a sufficiently smooth boundary $S_1$. The
scattering problem consists of finding the solution to the problem:
\be\label{e1} (\nabla^2+k^2)u=0\text{  in  } D'_1:=\R^3\setminus
D_1, \ee
\be\label{e2} \Gamma u=0\text{  on }S_1, \ee \be\label{e3}
u=u_0+v, \ee where \be\label{e4} u_0=e^{ik\alpha\cdot x}, \quad
\alpha\in S^2, \ee $S^2$ is the unit sphere in $\R^3$, $u_0$ is the
incident field, $v$ is the scattered field satisfying the radiation
condition
\be\label{e5} v_r-ikv=o\left(\frac{1}{r}\right),\quad
r:=|x|\to \infty,\ v_r:=\frac{\partial v}{\partial r}, \ee $\Gamma
u$ is the boundary condition (bc) of one of the following types
\be\label{e6} \Gamma u=\Gamma_1u=u\quad \text{(Dirichlet bc)}, \ee
\be\label{e7} \Gamma u=\Gamma_2u=u_N-\zeta_1u,\quad \text{
Im}\zeta_1\leq 0,\ \text{( impedance bc)}, \ee where $\zeta_1$ is a
constant, $N$ is the unit normal to $S_1$, pointing out of $D_1$,
and \be\label{e8} \Gamma u=\Gamma_3u=u_N, \ \text{( Neumann bc)}. \ee
The transmission bc are defined in Section 5, see \eqref{eq1} there.

It
is well known (see,
e.g., \cite{R190}) that problem \eqref{e1}-\eqref{e3} has a unique
solution. We now assume that \be\label{e9} a:=0.5\,
\text{diam}D_1,\quad ka\ll 1, \ee and look for the solution to problem
\eqref{e1}-\eqref{e3} of the form \be\label{e10} u(x)=u_0(x)+\int_{S_1}
g(x,t)\sigma_1(t)dt,\quad g(x,y):=\frac{e^{ik|x-y|}}{4\pi |x-y|},
\ee where $dt$ is the element of the surface area of $S_1$. One can
prove that the unique solution to the scattering problem
\eqref{e1}-\eqref{e3} with any of the boundary conditions
\eqref{e6}-\eqref{e8} can be found in the form \eqref{e10}, and the
function $\sigma_1$ in equation \eqref{e10} is uniquely defined from the
boundary condition \eqref{e2}. The scattering amplitude
$A(\beta,\alpha)=A(\beta,\alpha,k)$ is defined by the formula
\be\label{e11} v=\frac{e^{ikr}}{r}A(\beta,\alpha,k)+o\left(
\frac{1}{r}\right),\quad r\to \infty,\ \beta:=\frac{x}{r}.\ee The
equations for finding $\sigma_1$ are: \be\label{e3_1}
\int_{S_1}g(s,t)\sigma_1(t)dt=-u_0(s), \ee \be\label{e3_2}
u_{0N}-\zeta_1u_0+\frac{A\sigma_1-\sigma_1}{2}-\zeta_1\int_{S_1}g(s,t)\sigma_1(t)dt=0,
\ee \be\label{e3_3} u_{0N}+\frac{A\sigma_1-\sigma_1}{2}=0, \ee
respectively, for conditions \eqref{e6}-\eqref{e8}. The operator $A$
is defined as follows: \be\label{e3_4}
A\sigma:=2\int_{S_1}\frac{\partial}{\partial N_s}g(s,t)\sigma_1(t)dt.\ee
Equations \eqref{e3_1}-\eqref{e3_3} are uniquely solvable, but there are
no analytic formulas for their solutions for bodies of arbitrary
shapes. However, if the body $D_1$ is small, $ka\ll 1$, one can
rewrite \eqref{e10} as \be\label{e12}
u(x)=u_0(x)+g(x,0)Q_1+\int_{S_1}[g(x,t)-g(x,0)]\sigma_1(t)dt, \ee
where \be\label{e13} Q_1:=\int_{S_1}\sigma_1(t)dt, \ee and $0\in
D_1$ is the origin.

If $ka\ll1$, then we prove that \be\label{e14} |g(x,0)Q_1|\gg
\left|\int_{S_1}[g(x,t)-g(x,0)]\sigma_1(t)dt\right|,\quad |x|>a. \ee
Therefore, the scattered field is determined outside $D_1$
by a single number $Q_1$.
This number can be obtained analytically
without solving equations \eqref{e3_1}-\eqref{e3_2}. The case
\eqref{e3_3} requires a special approach by the reason discussed in
detail later.

Let us give the results for equations \eqref{e3_1} and
\eqref{e3_2} first. For equation \eqref{e3_1} one has \be\label{e15}
Q_1=\int_{S_1}\sigma_1(t)dt=-Cu_0(0)[1+o(1)],\quad a\to 0, \ee where
$C$ is the electric capacitance of a perfect conductor with the
shape $D_1$. For equation \eqref{e3_2} one has \be\label{e16}
Q_1=-\zeta |S_1|u_0(0)[1+o(1)],\quad a\to 0, \ee where $|S_1|$ is
the surface area of $S_1$. The scattering amplitude for problem
\eqref{e1}-\eqref{e3} with $\Gamma=\Gamma_1$ (acoustically soft
particle) is \be\label{e17}
A_1(\beta,\alpha)=-\frac{C}{4\pi}[1+o(1)], \ee since
$$u_0(0)=e^{ik\alpha\cdot x}|_{x=0}=1.$$ {\it Therefore,  in this case
the
scattering is isotropic and of the order $O(a)$, because the
capacitance $C=O(a)$. }

The scattering amplitude for problem
\eqref{e1}-\eqref{e3} with $\Gamma=\Gamma_2$ (small impedance
particles) is : \be\label{e18}
A_2(\alpha,\beta)=-\frac{\zeta_1|S_1|}{4\pi}[1+o(1)], \ee since
$u_0(0)=1$.

{\it In this case the scattering is also isotropic, and of
the order $O(\zeta|S_1|)$.}

If $\zeta_1=O(1)$, then $A_2=O(a^2)$,
because $|S_1|=O(a^2)$. If
$\zeta_1=O\left(\frac{1}{a^\kappa}\right)$, $\kappa\in(0,1)$, then
 $A_2=O(a^{2-\kappa})$. The case $\kappa=1$ was considered in \cite{R509}.

The scattering amplitude for problem
\eqref{e1}-\eqref{e3} with $\Gamma=\Gamma_3$ (acoustically hard
particles) is \be\label{e19}
A_3(\beta,\alpha)=-\frac{k^2|D_1|}{4\pi}(1+\beta_{pq}\beta_p\alpha_q),\,\,\text{
if } u_0=e^{ik\alpha\cdot x}. \ee
Here and below summation is understood over
the repeated indices, $\alpha_q=\alpha \cdot e_q$, $\alpha \cdot e_q$
denotes the
dot product of two vectors in $\R^3$, $p,q=1,2,3$,
$\{e_p\}$ is an orthonormal Cartesian basis of $\R^3$,
$|D_1|$ is the volume of $D_1$, $\beta_{pq}$
is the magnetic polarizability tensor defined as follows
(\cite{R476}, p.62): \be\label{e20}
\beta_{pq}:=\frac{1}{|D_1|}\int_{S_1}t_p\sigma_{1q}(t)dt, \ee
 $\sigma_{1q}$ is the solution to
the equation \be \label{e21} \sigma_{1q}(s)=A_0\sigma_{1q}-2N_q(s),
\ee  $N_q(s)=N(s)\cdot e_q$, $N=N(s)$ is the unit outer normal to $S_1$ at
the point $s$, i.e., the normal pointing out of $D_1$, and $A_0$
is the operator $A$ at $k=0$. For small bodies $\|A-A_0\|=o(ka)$.

If $u_0(x)$ is an arbitrary field satisfying equation \eqref{e1},
not necessarily the plane wave $e^{ik\alpha\cdot x}$, then
\be\label{e22}
A_3(\beta,\alpha)=\frac{|D_1|}{4\pi}\left(ik\beta_{pq}\frac{\partial
u_0}{\partial x_q} \beta_p+\triangle u_0\right). \ee The
above formulas are derived in Section 2. In Section 3 we develop a
theory for many-body wave scattering problem and derive the
equations for effective field in the medium, in which many small
particles are embedded, as $a\to 0$.

The results, presented in this paper, are based on the earlier works of
the author (\cite{R508}-\cite{R597}). Our presentation and some of the
results are novel. These results and methods of their derivation
differ much from those in the
homogenization theory (\cite{JKO}, \cite{MK}). The
differences are:

i) no periodic structure in the problems is assumed,

ii) the operators in our problems are non-selfadjoint and have
continuous spectrum,

iii) the limiting medium is not homogeneous and its
parameters are not periodic,

iv) the technique for passing to the limit
is different from one used in homogenization theory.

Let us summarize the results for one-body wave scattering.

{\bf Theorem 1.1} {\it The scattering amplitude for the problem
\eqref{e1}-\eqref{e4} for small body of an arbitrary shape are
given by formulas \eqref{e21}, \eqref{e22}, \eqref{e23},  
for the boundary conditions
$\Gamma_1$, $\Gamma_2$, $\Gamma_3$, respectively.}

\section{Derivation of the formulas for one-body wave scattering
problems}

Let us recall the known result (see e.g., \cite{R190})
\be\label{e23} \frac{\partial }{\partial
N_s^-}\int_{S_1}g(x,t)\sigma_1(t)dt=\frac{A\sigma_1-\sigma_1}{2}\ee
concerning the limiting value of the normal derivative of single-layer
potential from outside. Let $x_m\in D_m$, $t\in S_m$, $S_m$ is the
surface of $D_m$, $a=0.5\,\text{diam} D_m.$

In this Section $m=1$, and $x_m=0$ is the origin.

We assume  that $ka\ll 1$, $ad^{-1}\ll 1$,
so $|x-x_m|=d\gg a$. Then \be\label{e24}
\frac{e^{ik|x-t|}}{4\pi|x-t|}=\frac{e^{ik|x-x_m|}}{4\pi|x-x_m|}
e^{-ik(x-x_m)^o\cdot(t-x_m)}\left(1+O(ka+\frac{a}{d})\right),
\ee \be\label{e25}
k|x-t|=k|x-x_m|-k(x-x_m)^o\cdot(t-x_m)+O\left(\frac{ka^2}{d}\right),
\ee where
$$d=|x-x_m|,\quad (x-x_m)^o:=\frac{x-x_m}{|x-x_m|},$$ and
\be\label{e26} \frac{|x-t|}{|x-x_m|}=1+O\left(\frac{a}{d}\right).
\ee
Let us  derive estimate \eqref{e15}. Since
$|t|\leq a$ on $S_1$, one has
$$g(s,t)=g_0(s,t)(1+O(ka)),$$ where
$g_0(s,t)=\frac{1}{4\pi|s-t|}$. Since $u_0(s)$ is a smooth function,
one has $|u_0(s)-u_0(0)|=O(a)$. Consequently, equation \eqref{e3_1}
can be considered as an equation for electrostatic charge
distribution $\sigma_1(t)$ on the surface $S_1$ of a perfect
conductor $D_1$, charged to the constant potential $-u_0(0)$ (up to
a small term of the order $O(ka)$). It is known that the total charge
$Q_1=\int_{S_1}\sigma_1(t)dt$ of this conductor is equal to
\be\label{e27} Q_1=-Cu_0(0)(1+O(ka)), \ee where $C$ is the electric
capacitance of the perfect conductor with the shape $D_1$.

Analytic
formulas for electric capacitance $C$ of a perfect conductor of an
arbitrary shape, which allow to calculate $C$ with a desired
accuracy, are derived in \cite{R476}. For example, the zeroth
approximation formula is \be\label{e28}
C^{(0)}=\frac{4\pi|S_1|^2}{\int_{S_1}\int_{S_1}\frac{dsdt}{r_{st}}},\quad
r_{st}=|t-s|, \ee and we assume in \eqref{e28} that $\epsilon_0=1$,
where $\epsilon_0$ is the dielectric constant of the
homogeneous medium in which the
perfect conductor is placed. Formula
\eqref{e27} is formula \eqref{e15}. If $u_0(x)=e^{ik\alpha\cdot x}$,
then $u_0(0)=1$, and $Q_1=-C(1+O(ka))$. In this case
$$A_1(\beta,\alpha)=\frac{Q_1}{4\pi}=-\frac{C}{4\pi}[1+O(ka)],$$ which
is formula \eqref{e17}.

{\it Consider now wave scattering by an impedance particle}.

Let us derive formula \eqref{e16}. Integrate equation \eqref{e3_2}
over $S_1$, use the divergence formula \be\label{e29}
\int_{S_1}u_{0N} ds=\int_{D_1}\nabla^2u_0 dx=-k^2\int_{D_1}u_0
dx=k^2|D_1|u_0(0)[1+o(1)], \ee where $|D_1|=O(a^3)$, and the formula
\be\label{e30} -\zeta_1\int_{S_1}u_0ds=-\zeta_1|S_1|u_0(0)[1+o(1)].
\ee Futhermore $|\int_{S_1}g(s,t)ds|=O(a),$ so \be\label{e31}
\zeta_1\int_{S_1}ds\int_{S_1}g(s,t)\sigma_1(t)dt=O(aQ_1). \ee
Therefore, the term \eqref{e31} is negligible compared with $Q_1$
as  $a\to 0$. Finally, if $ka\ll1$, then
$g(s,t)=g_0(s,t)\left( 1+ik|s-t|+\hdots \right),$ and
\be\label{e32}
\frac{\partial}{\partial N_s}g(s,t)=\frac{\partial }{\partial
N_s}g_0(s,t)[1+O(ka)]. \ee Denote by $A_0$ the operator
\be\label{e33} A_0\sigma=2\int_{S_1}\frac{\partial
g_0(s,t)}{\partial N_s}\sigma_1(t)dt. \ee It is known from the
potential theory that
\be\label{e34} \int_{S_1}A_0\sigma_1
ds=-\int_{S_1}\sigma_1(t)dt,\quad 2\int_{S_1} \frac{\partial
g_0(s,t)}{\partial N_s}ds=-1,\quad t\in S_1.\ee Therefore,
\be\label{e35}
\int_{S_1}ds\frac{A\sigma_1-\sigma_1}{2}=-Q_1[1+O(ka)]. \ee
Consequently, from fromulas \eqref{e29}-\eqref{e35} one gets formula
\eqref{e18}.

One can see that the wave scattering by an impedance particle is
isotropic, and the scattered field is of the order
$O(\zeta_1|S_1|)$. Since $|S_1|=O(a^2)$, one would have
$O(\zeta_1|S_1|)=O(a^{2-\kappa})$ if
$\zeta_1=O\left(\frac{1}{a^\kappa}\right)$, $\kappa\in(0,1)$.

{\it Consider now wave scattering by an acoustically hard small particle,
i.e., the problem with the Neumann boundary condition.}

In this case
we will prove that:

i)  The scattering is
anisotropic,

ii) It is defined not by a single number, as in the previous
two cases, but by a tensor,

and

iii) The order of the scattered field is
$O(a^3)$ as $a\to 0$, for a fixed $k>0$, i.e., the scattered field
is much smaller than in the previous two cases.

When one integrates
over $S_1$ equation \eqref{e3_2}, one gets \be\label{e36}
Q_1=\int_{D_1}\nabla^2u_0 dx=\nabla^2u_0(0)|D_1|[1+o(1)],\quad a\to
0. \ee Thus, $Q_1=O(a^3)$. Therefore, the contribution of the term
$e^{-ikx^o\cdot t}$ in formula \eqref{e24} with $x_m=0$ will be also
of the order $O(a^3)$ and should be taken into account, {\it in contrast
to the previous two cases.} Namely, \be\label{e37}
u(x)=u_0(x)+g(x,0)\int_{S_1}e^{-ik\beta\cdot t}\sigma_1(t)dt,\quad
\beta:=\frac{x}{|x|}=x^o. \ee One has \be\label{e38}
\int_{S_1}e^{-ik\beta\cdot
t}\sigma_1(t)dt=Q_1-ik\beta_p\int_{S_1}t_p\sigma_1(t)dt, \ee where
the terms of higher order of smallness are neglected and
summation over index $p$ is understood. The function $\sigma_1$
solves equation \eqref{e3_3}: \be\label{e39}
\sigma_1=A\sigma_1+2u_{0N}=A\sigma_1+2ik\alpha_qN_qu_0(s),\quad s\in
S_1 \ee if $u_0(x)=e^{ik\alpha\cdot x}$.

Comparing \eqref{e39} with
\eqref{e21}, using \eqref{e20}, and taking into account that $ka\ll
1$, one gets \be\label{e40}\begin{split}
-ik\beta_p\int_{S_1}t_p\sigma_1(t)dt&=-ik\beta_p|D_1|\beta_{pq}(-ik\alpha_q)
u_0(0)[1+O(ka)]\\
&=-k^2|D_1|\beta_{pq}\beta_p\alpha_qu_0(0)[1+O(ka)].
\end{split}\ee
From \eqref{e36}, \eqref{e38} and \eqref{e40} one gets formula
\eqref{e19}, because $\nabla^2u_0=-k^2u_0.$

If $u_0(x)$ is an
arbitrary function, satisfying equation \eqref{e1}, then
$ik\alpha_q$ in \eqref{e39} is replaced by $\frac{\partial
u_0}{\partial x_q}$, and $-k^2u_0=\triangle u_0$, which yields
formula \eqref{e22}.

This completes the derivation of the formulas for the solution of
scalar wave scattering problem by one small body on the boundary of
which the Dirichlet, or the impedance, or the Neumann boundary
condition is imposed.

\section{Many-body scattering problem}
In this Section we assume that there are $M=M(a)$ small bodies
(particles) $D_m$, $1\leq m\leq M$, $a=0.5\max\text{diam}D_m$, $ka\ll1$.
The
distance $d=d(a)$ between neighboring bodies is much larger than
$a$, $d\gg a$, but we do not assume that $d\gg \lambda$, so {\it there
may be many small particles on the distances of the order of the
wavelength $\lambda$.} This means that our medium with the embedded
particles is not necessarily diluted.

We assume that the small bodies are embedded in an arbitrary large but
finite domain $D$, $D\subset \R^3$, so
$D_m\subset D$. Denote
$D':=\R^3\setminus D$ and $\Omega:=\cup_{m=1}^M D_m,$
$S_m:=\partial
D_m$, $\partial \Omega=\cup_{m=1}^M S_m$. By $N$ we denote a unit
normal to $\partial \Omega$, pointing out of $\Omega$, by $|D_m|$
the volume of the body $D_m$ is denoted.

The
scattering problem consists of finding the solution to the following
problem \be\label{e41} (\nabla^2+k^2)u=0\text{  in  } \R^3\setminus
\Omega, \ee \be\label{e42} \Gamma u=0\text{   on  } \partial \Omega,
\ee \be\label{e43} u=u_0+v, \ee where $u_0$ is the incident field,
satisfying equation \eqref{e41} in $\R^3$, for example,
$u_0=e^{ik\alpha \cdot x}$, $\alpha\in S^2$, and $v$ is the
scattered field, satisfying the radiation condition \eqref{e5}. The
boundary condition \eqref{e42} can be of the types
\eqref{e6}-\eqref{e8}.

In the case of impedance boundary condition \eqref{e7} we assume
that \be\label{e44} u_N=\zeta_mu\text{ on } S_m,\quad 1\leq m\leq
M,\ee so the impedance may vary from one particle to another. We
assume that \be\label{e45} \zeta_m=\frac{h(x_m)}{a^\kappa},\quad
\kappa\in(0,1), \ee where $x_m\in D_m$ is a point in $D_m$, and
$h(x),$ $x\in D$, is a given function, which we can choose as we
wish, subject to the condition  Im$h(x)\leq 0$. For simplicity we
assume that $h(x)$ is a
continuous function.

Let us make the following assumption about the
distribution of small particles: if $\Delta\subset D$ is an
arbitrary open subset of $D$, then the number $\mathcal{N}(\Delta)$
of small particles in $\Delta$, assuming the impedance boundary condition,
is: \be\label{e46}
\mathcal{N}_\zeta(\Delta)=\frac{1}{a^{2-\kappa}}\int_{\Delta}N(x)dx[1+o(1)],
\quad a\to 0, \ee where $N(x)\geq 0$ is a given function.
If the
Dirichlet boundary condition is assumed, then \be\label{e47}
\mathcal{N}_D(\Delta)=\frac{1}{a}\int_{\Delta}N(x)dx[1+o(1)],\quad
a\to 0.\ee The case of the Neumann boundary condition will be
considered later.

We look for the solution to problem
\eqref{e41}-\eqref{e43} with the Dirichlet boundary condition of the
form \be\label{e48}
u=u_0+\sum_{m=1}^M\int_{S_m}g(x,t)\sigma_m(t)dt,\ee where
$\sigma_m(t)$ are some functions to be determined from the boundary
condition \eqref{e42}. It is proved in \cite{R509} that problem
\eqref{e41}-\eqref{e43} has a unique solution of the form
\eqref{e48}. For any $\sigma_m(t)$ function \eqref{e48} solves
equation \eqref{e41} and satisfies condition \eqref{e43}. The
boundary condition \eqref{e42} determines $\sigma_m$ uniquely.
However, if $M\gg1$, then numerical solution of the system of
integral equations for $\sigma_m$, $1\leq m\leq M$, which one gets
from the boundary condition \eqref{e42}, is practically not
feasible.

{\it To avoid this principal difficulty we prove that the
solution to scattering problem \eqref{e41}-\eqref{e43} is determined
by $M$ numbers \be\label{e49} Q_m:=\int_{S_m}\sigma_m(t)dt, \ee
rather than $M$ functions $\sigma_m(t)$.}

This is possible to prove
if the
particles $D_m$ are small. We derive analytical formulas for $Q_m$
as $a\to 0$.

Let us define the effective (self-consistent) field
$u_e(x)=u_e^{(j)}(x)$, acting on the $j-$th particle, by the formula
\be\label{e50} u_e(x):=u(x)-\int_{S_j}g(x,t)\sigma_j(t)dt,\quad
|x-x_j|\sim a.\ee Physically this field acts on the $j-$th particle
and is a sum of the incident field and the fields acting from all
other particles: \be\label{e51}
u_e(x)=u_e^{(j)}(x):=u_0(x)+\sum_{m\neq
j}\int_{S_m}g(x,t)\sigma_m(t)dt. \ee Let us rewrite \eqref{e51} as
follows: \be\label{e52} u_e(x)=u_0(x)+\sum_{m\neq j}^M
g(x,x_m)Q_m+\sum_{m\neq
j}^M\int_{S_m}[g(x,t)-g(x,x_m)]\sigma_m(t)dt. \ee
We want to prove
that the last sum is negligible compared with the first one as $a\to
0$. To prove this, let us give some estimates. One has $|t-x_m|\leq a$,
$d=|x-x_m|$, \be\label{e53}
|g(x,t)-g(x,x_m)|=\max\left\{
O\left(\frac{a}{d^2}\right),O\left(\frac{ka}{d}\right)\right\},\quad
|g(x,x_m)|=O(1/d). \ee Therefore, if $|x-x_j|=O(a)$, then
\be\label{e54}
\frac{\left|\int_{S_m}[g(x,t)-g(x,x_m)]\sigma_m(t)dt\right|}{|g(x,x_m)Q_m|}\leq
O(ad^{-1}+ka). \ee One can also prove that \be\label{e55}
J_1/J_2=O(ka+ad^{-1}),\ee where $J_1$ is the first sum in
\eqref{e52} and $J_2$ is the second sum in \eqref{e52}. Therefore,
at any point $x\in \Omega'=\R^3\setminus \Omega$ one has
\be\label{e56} u_e(x)=u_0(x)+\sum_{m=1}^M g(x,x_m)Q_m,\quad x\in
\Omega', \ee
where the terms of higher order of smallness are omitted.

\subsection{ The case of acoustically soft particles}
If \eqref{e42} is the Dirichlet condition, then, as we have proved
in Section 2  (see formula \eqref{e27}), one has
\be\label{e57} Q_m=-C_mu_e(x_m). \ee Thus, \be\label{e58}
u_e(x)=u_0(x)-\sum_{m=1}^M g(x,x_m)C_mu_e(x_m), \quad x\in \Omega'.\ee
One has \be\label{e59}
u(x)=u_e(x)+o(1),\quad a\to 0, \ee so the full field and
effective field are practically the same.

Let us write a linear
algebraic system (LAS) for finding unknown quantities $u_e(x_m)$:
\be\label{e60} u_e(x_j)=u_0(x_j)-\sum_{m\neq
j}^Mg(x_j,x_m)C_mu_e(x_m). \ee If $M$ is not very large, say
$M=O(10^3)$, then LAS \eqref{e60} can be solved numerically, and
formula \eqref{e58} can be used for calculation of $u_e(x)$.

Consider the limiting case, when $a\to 0$. One can rewrite
\eqref{e60} as follows: \be\label{e61}
u_e(\xi_q)=u_0(\xi_q)-\sum_{p\neq q}^P
g(\xi_q,\xi_p)u_e(\xi_p)\sum_{x_m\in \Delta_p}C_m, \ee where
$\{\Delta_p\}_{p=1}^P$ is a union of cubes which forms a covering of
$D$,
$$\max_p diam \Delta_p:=b=b(a)\gg a,$$
 \be\label{e62}
\lim_{a\to 0}b(a)=0. \ee By $|\Delta_p|$ we denote the volume
(measure) of $\Delta_p$, and $\xi_p$ is the center of $\Delta_p$,
or a point $x_p$ in an arbitrary small body $D_p$, located in $\Delta_p$.
Let us assume that there exists the limit \be\label{e63} \lim_{a\to
0}\frac{\sum_{x_m\in \Delta_p}C_m}{|\Delta_p|}=C(\xi_p),\quad
\xi_p\in \Delta_p. \ee
For example, one may have  \be\label{e64} C_m=c(\xi_p)a
\ee for all $m$ such that $x_m\in \Delta_p$, where $c(x)$ is some
function in $D$. If all $D_m$ are balls of radius $a$,
then $c(x)=4\pi$. We have \be\label{e65}
\sum_{x_m\in\Delta_p}C_m=C_pa\mathcal{N}(\Delta_p)=C_pN(\xi_p)|\Delta_p|[1+o(1)],\quad
a\to 0, \ee so limit \eqref{e63} exists, and \be\label{e66}
C(\xi_p)=c(\xi_p)N(\xi_p).\ee From \eqref{e61},
\eqref{e64}-\eqref{e66} one gets \be\label{e67}
u_e(\xi_q)=u_0(\xi_q)-\sum_{p\neq
q}g(\xi_q,\xi_p)c(\xi_p)N(\xi_p)u_e(\xi_p)|\Delta_p|,\quad 1\leq
p\leq P. \ee Linear algebraic system  \eqref{e67} can be considered as the
{\it collocation
method for solving integral equation} \be\label{e68}
u(x)=u_0(x)-\int_Dg(x,y)c(y)N(y)u(y)dy. \ee It is proved in
\cite{R573} that system \eqref{e67} is uniquely solvable for all
sufficiently small $b(a)$, and the function \be\label{e69}
u_P(x):=\sum_{p=1}^P\chi_p(x)u_e(\xi_p)\ee converges in $L^\infty
(D)$ to the unique solution of equation \eqref{e68}.
The function $\chi_p(x)$ in \eqref{e69} is the characteristic function
of the cube $\Delta_p$: it is equal to $1$ in $\Delta_p$ and vanishes
outside $\Delta_p$.
Thus, if $a\to
0$, the solution to the many-body wave scattering problem in the
case of the Dirichlet boundary condition is well approximated by the
unique solution of the integral equation \eqref{e68}.

Applying the
operator $L_0:=\nabla^2+k^2$ to \eqref{e68}, and using the formula
$L_0g(x,y)=-\delta(x-y)$, where $\dl(x)$ is the delta-function, one
gets \be\label{e70}\nabla^2u+k^2u-q(x)u=0 \text{ in } \R^3,\quad
q(x):=c(x)N(x).\ee
The physical conclusion is:

{\it If one  embeds $M(a)=O(1/a)$ small
acoustically
soft particles, which are distributed as in \eqref{e47}, then one
creats, as $a\to 0$, a
limiting medium, which is inhomogeneous,  and has a refraction coefficient
$n^2(x)=1-k^{-2}q(x).$ }

It is interesting from the physical point of
view to note that {\it the limit, as $a\to 0$, of the total volume of the
embedded particles is zero.}

Indeed, the volume of one particle is
$O(a^3)$, the total number $M$ of the embedded particles is
$O(a^3M)=O(a^2)$, and $\lim_{a\to 0}O(a^2)=0$.

The second
observation is: if \eqref{e47} holds, then on a unit length
straight line there are $O(\frac{1}{a^{1/3}})$ particles, so the
distance between neighboring particles is $d=O(a^{1/3})$. If
$d=O(a^\gamma)$ with $\gamma>\frac{1}{3}$, then the number of the
embedded particles in a subdomain $\Delta_p$ is
$O(\frac{1}{d^3})=O(a^{-3\gamma})$. In this case, for $3\gamma>1$,
the limit in \eqref{e65} is $C(\xi_p)=\lim_{a\to
0}c_paO(a^{-3\gamma})=\infty$. Therefore, the product of this limit
by $u$ remains finite only if $u=0$ in $D$. Physically this means
that if the distances between neighboring perfectly soft particles
are smaller than $O(a^{1/3})$, namely, they are $O(a^\gamma)$ with
any $\gamma>\frac{1}{3}$, then $u=0$ in $D$.

On the other hand, if
$\gamma<\frac{1}{3}$, then the limit $C(\xi_p)=0$, and $u=u_0$ in
$D$, so that the embedded particles do not change, in the limit
$a\to 0$, properties of the medium.

This concludes our discussion of
the scattering problem for many acoustically soft particles.

\subsection{ Wave scattering by many impedance particles}
We assume now that \eqref{e45} and \eqref{e46} hold, use the exact
boundary condition \eqref{e42} with $\Gamma=\Gamma_2$, that is,
\be\label{e71}
u_{eN}-\zeta_mu_e+\frac{A_m\sigma_m-\sigma_m}{2}-\zeta_m\int_{S_m}g(s,t)\sigma_m(t)dt=0,
\ee and integrate \eqref{e71} over $S_m$ in order to derive an
analytical asymptotic formula for $Q_m=\int_{S_m}\sigma_m(t)dt.$

We have
\be\label{e72} \int_{S_m}u_{eN}ds=\int_{D_m}\nabla^2 u_e
dx=O(a^3), \ee \be\label{e73}
\int_{S_m}\zeta_mu_e(s)ds=h(x_m)a^{-\kappa}|S_m|u_e(x_m)[1+o(1)],\quad
a\to 0, \ee \be\label{e74}
\int_{S_m}\frac{A_m\sigma_m-\sigma_m}{2}ds=-Q_m[1+o(1)],\quad a\to
0, \ee and \be\label{e75}
\zeta_m\int_{S_m}\int_{S_m}g(s,t)\sigma_m(t)dt=h(x_m)a^{1-\kappa}Q_m=o(Q_m),\quad
0<\kappa<1. \ee From \eqref{e71}-\eqref{e75} one finds
\be\label{e76} Q_m=-h(x_m)a^{2-\kappa}|S_m|a^{-2}u_e(x_m)[1+o(1)].
\ee This yields the formula for the approximate solution to the
wave scattering problem for many impedance particles:
\be\label{e77}
u(x)=u_0(x)-a^{2-\kappa}\sum_{m=1}^Mg(x,x_m)b_mh(x_m)u_e(x_m)[1+o(1)],
\ee where $$b_m:=|S_m|a^{-2}$$ are some positive numbers which depend
on the geometry of $S_m$ and are independent of $a$. For example, if
all $D_m$ are balls of radius $a$, then $b_m=4\pi$.

{\it A linear algebraic system for $u_e(x_m)$}, analogous to \eqref{e60},
is \be\label{e78} u_e(x_j)=u_0(x_j)-a^{2-\kappa}\sum_{m=1,m\neq j}^M
g(x_j,x_m)b_m h(x_m)u_e(x_m). \ee The integral equation for the
limiting effective field in the medium with embedded small
particles, as $a\to 0$, is \be\label{e79}
u(x)=u_0(x)-b\int_{D}g(x,y)N(y)h(y)u(y)dy, \ee where \be\label{e80}
u(x)=\lim_{a\to 0}u_e(x),\ee and we have assumed in \eqref{e79} for
simplicity that $b_m=b$ for all $m$, that is, all small particles
are of the same shape and size.

Applying operator $L_0=\nabla^2+k^2$
to equation \eqref{e79}, one finds the differential equation for the
limiting effective field $u(x)$: \be\label{e81}
(\nabla^2+k^2-bN(x)h(x))u=0\text{  in }\R^3, \ee and $u$ satisfies
condition \eqref{e43}.

{\it The conclusion is: the limiting medium is
inhomogeneous, and its properties are described by the function
\be\label{e82} q(x):=bN(x)h(x). \ee }
Since the choice of the
functions $N(x)\geq 0$ and $h(x)$, Im$h(x)\leq 0$, is at our
disposal, we can create the medium with desired properties by
embedding many small impedance particles, with suitable impedances,
according to the distribution law \eqref{e46} with a suitable
$N(x)$. The function \be\label{e83} 1-k^{-2}q(x)=n^2(x) \ee is the
refraction coefficient of the limiting medium. Given a desired
refraction coefficient $n^2(x)$, Im$n^2(x)\geq 0$, one can find
$N(x)$ and $h(x)$ so that \eqref{e83} holds, that is, one can create
a material with a desired refraction coefficient by embedding into a
given material many small particles with suitable boundary
impedances.

This concludes our discussion of the wave scattering
problem with many small impedance particles.
\subsection{Wave scattering by many  acoustically hard particles}
Consider now the case of acoustically hard particles, i.e., the case
of Neumann boundary condition. The exact boundary integral equation
for the function $\sigma_m$
in this case is: \be\label{e84}
u_{eN}+\frac{A_m\sigma_m-\sigma_m}{2}=0. \ee Arguing as in Section
2, see formulas \eqref{e36}-\eqref{e40}, one obtains \be\label{e85}
u_e(x)=u_0(x)+\sum_{m=1}^Mg(x,x_m)\left[\triangle
u_e(x_m)+ik\beta_{pq}^{(m)}\frac{(x_p-(x_m)_p)}{|x-x_m|}\frac{\partial
u_e(x_m)}{\partial (x)_q}\right]|D_m|. \ee Here we took into account
that the unit vector $\beta$ in \eqref{e40} is now the vector
$\frac{x-x_m}{|x-x_m|}$, and
$\beta_p=\frac{(x)_p-(x_m)_p}{|x-x_m|}$, where $(x)_p:=x\cdot e_p$
is the $p-$th component of vector $x$ in the Euclidean orthonormal basis
$\{e_p\}_{p=1}^3$.

There are three sets of unknowns in \eqref{e85}:
$u_e(x_m)$, $\frac{\partial u_e(x_m)}{\partial (x)_q}$, and
$\triangle u_e(x_m)$, $1\leq m\leq M$, $1\leq q\leq 3$. To obtain
linear algebraic system for $u_e(x_m)$ and $\frac{\partial
u_e(x_m)}{\partial (x)_q}$ one sets $x=x_j$ in \eqref{e85}, takes
the sum in \eqref{e85} with $m\neq j$. This yields the first set of
equations for finding these unknowns. Then one takes derivative
of equation \eqref{e85} with respect to $(x)_q$, sets $x=x_j$,
and takes the sum in \eqref{e85} with $m\neq j$. This yields the second
set of equations for finding these unknowns. Finally, one takes
Laplacian of equation \eqref{e85}, sets $x=x_j$, and takes the sum in
\eqref{e85} with $m\neq j$. This yields the third set of linear
algebraic equations
for finding $u_e(x_m)$, $\frac{\partial u_e(x_m)}{\partial (x)_q}$,
and $\Delta u_e(x_m)$.

Passing to the limit $a\to 0$
in equation \eqref{e85}, yields the equation for the limiting field
\be\label{e86}
u(x)=u_0(x)+\int_Dg(x,y)\left(\rho(y)\nabla^2u(y)+ik\frac{\partial
u (y)}{\partial y_q}\frac{x_p-y_p}{|x-y|}B_{pq}(y)\right)dy, \ee where
$\rho(y)$ and $B_{pq}(y)$ are defined below, see formulas \eqref{e88}
and \eqref{e89}.

Let us derive equation \eqref{e86}.
We start by transforming the sum in \eqref{e85}. Let
$\{\Delta_l\}_{l=1}^L$ be a covering of $D$ by cubes $\Delta_l$,
$\max_l$ diam$\Delta_l=b=b(a)$. We assume that
$$b(a)\gg d\gg a, \qquad \lim_{a\to
0}b(a)=0.$$
Thus, there are many small particles $D_m$ in $\Delta_l$.
Let $x_l$ be a point in $\Delta_l$. One has
\be\label{e87}\begin{split} &\sum_{m=1}^Mg(x,x_m)\left[\triangle
u_e(x_m)+ik\frac{\partial u_e(x_m)}{\partial
(x)_q}\beta_{pq}^{(m)}\frac{((x)_p-(x_m)_p)}{|x-x_m|}\right]|D_m|\\
&=\sum_{l=1}^Lg(x,x_l)\left[\triangle u_e(x_l)\sum_{x_m\in
\Delta_l}|D_m|+ik\frac{\partial u_e(x_l)}{\partial
(x)_q}\frac{((x)_p-(x_l)_p)}{|x-x_l|}\sum_{x_m\in
\Delta_l}\beta_{pq}^{(m)}|D_m|\right].\end{split}\ee Assume that the
following limit exist: \be\label{e88} \lim_{a\to 0,y\in
\Delta_l}\frac{\sum_{x_m\in \Delta_l} |D_m|}{|\Delta_l|}=\rho(y),
\ee \be\label{e89} \lim_{a\to 0,y\in \Delta_l}\frac{\sum_{x_m\in
\Delta_l} \beta_{pq}^{(m)}|D_m|}{|\Delta_l|}=B_{pq}(y), \ee and
\be\label{e90} \lim_{a\to 0}u_e(y)=u(y),\quad \lim_{a\to
0}\frac{\partial u_e(y)}{\partial (y)_q}=\frac{\partial u (y)}{\partial
y_q},\quad \lim_{a\to 0}\nabla^2 u_e(y)=\nabla^2 u(y). \ee
 Then, the sum in \eqref{e87} converges to
\be\label{e91} \int_Dg(x,y)\left(\rho(y)\nabla^2u(y)+ik\frac{\partial
u(y)}{\partial y_q}\frac{x_p-y_p}{|x-y|}B_{pq}(y)\right)dy. \ee
Consequently, \eqref{e85} yields in the limit $a\to 0$ equation
\eqref{e86}. Equation \eqref{e86} cannot be reduced to a
differential equation for $u(x)$, because \eqref{e86} is an
integrodifferential equation whose integrand depends on $x$ and $y$.

Let us summarize the results in the following theorem.

{\bf Theorem 3.1.} {\it The many-body scattering problem
\eqref{e41}-\eqref{e43} has a unique solution for the Dirichlet, 
impedance, and Neumann boundary conditions. The limiting effective fields 
in the medium obtained by embedding many small particles of an arbitrary 
shape satisfy the equations (74), (85), and (90), for the Dirichlet, 
impedance, and Neumann boundary condiions, respectively.}

\section{Scattering by small particles embedded in an inhomogeneous
medium}
Suppose that the operator $\nabla^2+k^2$ in \eqref{e1} and
in \eqref{e41} is replaced by the operator
$L_0=\nabla^2+k^2n_0^2(x)$, where $n_0^2(x)$ is a known function,
\be\label{e92} \text{Im}\,n_0^2(x)\geq 0. \ee The function $n_0^2(x)$
is the refraction coefficient of an inhomogeneous medium in which
many small particles are embedded. The results, presented in Section
1-3 remain valid if one replaces function $g(x,y)$ by the Green's
function $G(x,y)$, \be\label{e93}
[\nabla^2+k^2n_0^2(x)]G(x,y)=-\delta(x-y),\ee satisfying the
radiation condition. We assume that \be\label{e94} n_0^2(x)=1\text{
in }D':=\R^3\setminus D. \ee The function $G(x,y)$ is uniquely
defined (see, e.g., \cite{R509}). The derivations of the results
remain essentially the same because \be\label{e95}
G(x,y)=g_0(x,y)[1+O(|x-y|)],\quad |x-y|\to 0, \ee where
$g_0(x,y)=\frac{1}{4\pi|x-y|}$. Estimates of $G(x,y)$ as $|x-y|\to
0$ and as $|x-y|\to \infty$ are obtained in \cite{R509}. Smallness
of particles in an inhomogeneous medium with refraction coefficient
$n_0^2(x)$ is described by the relation $kn_0a\ll 1$, where
$n_0:=\max_{x\in D}|n_0(x)|$, and $a=\max_{1\leq m\leq M}$diam$D_m$.

\section{Wave scattering by small bodies with
transmission (interface) boundary conditions}

There is a large literature on "homogenization", which deals with
the properties of the medium in which other materials is
distributed. Quite often it is assumed that the medium is periodic,
and homogenization is considered in the framework of G-convergence
(\cite{JKO},\cite{MK}). In most cases, one considers elliptic or
parabolic problems with elliptic operators positive-difinite and
having discrete spectrum.

A theory of wave scattering by many small
particles embedded in an inhomogeneous medium has been developed by the
author
(\cite{R509}-\cite{R607}). One of the pratically important
consequences of his theory was a derivation of the equation for the
effective (self-consistent) field in the limiting medium, obtained
in the limit $a \to 0$, $M=M(a) \to \infty$, where $a$ is the
characteristic size of a small particle, and $M(a)$ is the total
number of the embedded particles.

The theory was developed in Sections 1-4 (see
also papers \cite{R509}-\cite{R607}) for boundary
conditions (bc) on the surfaces of small bodies, which include the
Dirichlet bc, $u|_{S_m}=0$, where $S_m$ is the surface of the $m$-th
particle $D_m$, the impedance bc, $\zeta_m u|_{S_m}=u_N|_{S_m}$,
where $N$ is the unit normal to $S_m$, pointing out of $D_m$,
$\zeta_m$ is the boundary impedance, and the Neumann bc,
$u_N|_{S_m}=0$.

In the rest of this paper the development is presented of a similar
theory for
the  {\it transmission (interface)} bc:
\begin{equation} \label{eq1}
    \rho_m u_N^+ = u_N^-,\quad u^+=u^- \quad\text{ on } S_m, 1 \leq m
\leq M. \end{equation}
Here $\rho_m$ is a constant, +(-) denotes the limit of
$\frac{\partial u}{\partial N}$, from inside (outside) of $D_m$.

The physical meaning of the  transmission boundary conditions is the
continuity of the pressure and the normal component of the velocity
across the boundaries of the discontinuity of the density. One may
think about problem \eqref{eq1}-\eqref{eq5} (see below) as of the
problem of
acoustic wave scattering by many small bodies.

The essential novelty of the theory, developed in this paper, is the
asymptotically exact, as $a\to 0$,  treatment of the one-body  and
many-body scalar wave scattering problem in the case of small
scatterers on the boundaries of which the  transmission boundary
conditions are imposed. An analytic explicit asymptotic formula for
the field scattered  by one small body is derived. An integral
equation for the limiting effective field in the medium, in which
many small bodies are embedded, is derived in the limit $a\to 0$ and
$M(a)\to \infty$, where $M(a)$ is the total number of the embedded
small bodies (particles), and $M=M(a)$ tends to infinity at a
suitable rate as $a\to 0$.

For the problem with the number $M$ of particles not large, say,
less than 5000, our theory gives an efficient numerical method for
solving many-body wave scattering problem.

For the problem with  $M$ very large, say, larger than $10^5$, the
solution to many-body wave scattering problem consists in numerical
solution of the integral equation for the limiting field in the
medium, in which small particles are embedded. The solution to this
equation approximates the solution to the  many-body wave scattering
problem with high accuracy.

Our approach is quite different from the approach developed in
homogenization theory, we do not assume periodicity in the location
of the small scatterers. Our results are of interest also in the
case when the number of scatterers is not large, so the
homogenization theory is not applicable.

Let us formulate the scattering problem we are treating.
Below condition \eqref{eq1} is assumed. Let
\begin{align}
    &\text{Let } \Omega := \bigcup_{m=1}^M D_m, \quad
\Omega'=\mathbb{R}^3 \backslash \Omega, \nonumber \\
    &(\nabla^2 + k^2)u=0 \quad\text{ in } \Omega', \label{eq2} \\
    &(\nabla^2 + k_m^2)u=0 \quad\text{ in } D_m,  \quad 1 \leq m
\leq M, \label{eq3} \\
    &u=u_0 + v, u_0 = e^{ik \alpha \cdot x}, \quad \alpha \in S^2,
S^2 \text{ is a unit sphere in } \mathbb{R}^3, \label{eq4} \\
    &r\left(\frac{\partial v}{\partial r} - ikv \right)= o(1), \quad
r \to \infty. \label{eq5} \end{align} We assume that $\rho_m$, $k$ and
$k_m^2$ are fixed given positive constants, and the surfaces $S_m$
are smooth. A sufficient smoothness condition is $S_m \in
C^{1,\mu}$, $\mu \in (0,1)$, where $S_m$ in local coordinates is
given by a continuously differentiable function whose first
derivatives are H\"{o}lder-continuous with exponent $\mu$.

We assume that $x_m \in D_m$ is a point inside $D_m$, $a=\frac{1}{2}
\text{diam}D_m$, $d=O(a^\frac{1}{3})$ is the distance between the
neighboring particles, $\mathcal{N}(\Delta)=\sum_{x_m \in \Delta}1$,
is the number of particles in an arbitrary open set $\Delta$, the
domains $D_m$ are not intersecting, and
\begin{equation} \label{eq6}
    \mathcal{N}(\Delta)= \frac{1}{V} \int_{\Delta} N(x) dx[1+o(1)],
\quad a \to 0, \end{equation} where $N(x)\geq 0$ is a function which is
at our disposal, $V$ is the volume of one small body, $V=O(a^3)$. If
$D_m$ are balls of radius $a$, then $V=\frac{4\pi a^3}{3}$.

It is proved in \cite{R190} that problem \eqref{eq1}-\eqref{eq5} has
a unique solution.

We study wave scattering by a single small body in Section 6. In
other words, we study in Section 6 problem \eqref{eq1}-\eqref{eq5}
with $M=1$. The basic results of this Section are formulated in
Theorem 6.1.

In section 7 wave scattering by many small bodies  is considered.
The basic results of this Section are formulated in Theorem 7.1.
We always assume that
\begin{equation}
\label{eq7}
    ka << 1,\qquad d=O(a^\frac{1}{3}).
\end{equation}
\section{Wave scattering by one small body}
Let us look for the solution to problem \eqref{eq1}-\eqref{eq5} with
$M=1$ of the form
\begin{equation} \label{eq8}
    u(x)=u_0(x) + \int_S g(x,t)\sigma(t)dt+\varkappa\int_D g(x,y)u(y)dy,
\end{equation}
where $S=S_1$, $D=D_1$,
\begin{equation} \label{eq9}
    \varkappa: = k_1^2-k^2,\qquad
g(x,y):=\frac{e^{ik|x-y|}}{4\pi|x-y|},
\end{equation}
and $\sigma(t)$ is to be found so that conditions \eqref{eq1} are
satisfied. For any $\sigma \in C^{0,\mu_1}$, $\mu_1 \in (0,1]$,
where $C^{0,\mu_1}$ is the set of H\"older-continuous functions with
H\"older's exponent $\mu_1$,
 the solution to equation
\eqref{eq8} satisfies equations \eqref{eq2} and \eqref{eq3} with
$M=1$, and  equations \eqref{eq4} and \eqref{eq5}. This is easily
checked by a direct calculation. The second condition \eqref{eq1} is
also satisfied. To satisfy the first condition in  equations
\eqref{eq1} with $\rho_1=\rho$, one has to satisfy the following
equation
\begin{equation} \label{eq10}
    (\rho -1)u_{0_N}+\rho\frac{A\sigma+\sigma}{2}-\frac{A\sigma-\sigma}{2}+
(\rho-1)\frac{\partial}{\partial N_s}Bu=0,
\end{equation}
where
\begin{equation} \label{eq11}
    A\sigma=2\int_S \frac{\partial g(s,t)}{\partial N_S}
\sigma(t)dt, \quad Bu=\varkappa\int_D g(x,y)u(y)dy, \end{equation} and
the well-known formulas for the limiting values of the normal
derivatives of the single-layer potential $T\sigma:=\int_S
g(x,t)\sigma(t)dt$ on $S$ from inside and outside $D$ was used.

In \cite{R190} one finds a proof of the following existence and
uniqueness result. Let $H^2(D)$ denote the usual Sobolev space of
functions twice differentiable in $L^2$-sense.

{\bf Proposition 1.} {\it The system of equations \eqref{eq8} and
\eqref{eq10} for the unknown functions $\sigma$ on $S$ and $u(x)$ in
$D$ has a solution and this solution is unique in $C^{0,\mu_1}\times
H^2(D)$.}

If the solution $\{\sigma, u(x)|_{x\in D}\}$ is found, then formula
\eqref{eq8} defines $u=u(x)$ in $\mathbb{R}^3$.

Let us rewrite \eqref{eq10} as
\begin{equation} \label{eq12}
    \sigma=\lambda A \sigma +2\lambda B_1u + 2\lambda u_{0_N},
\end{equation}
where
\begin{equation} \label{eq13}
    \lambda = \frac{1-\rho}{1+\rho},
\quad B_1 u =\varkappa \frac{\partial}{\partial N_s} \int_D g(x,y)u(y)dy.
\end{equation}
If $\rho\in [0, \infty)$ then $\lambda\in (-1,1)$. Let us now use
the first assumption \eqref{eq7}, that is, the smallness of $a$. One
has:
\begin{align}
    &g(s,t)=g_0 (s,t)(1+O(ka)), \quad a \to 0; \quad
g_0(s,t)=\frac{1}{4 \pi |s-t|}, \label{eq14} \\
    &\frac{\partial}{\partial
N_s}\frac{e^{ik|s-t|}}{4\pi|s-t|}=\frac{\partial g_0}{\partial
N_s}(1+O((ka)^2)), \quad a\to 0, \label{eq15} \\
    &\text{so } A=A_0(1+O((ka)^2)), \quad a \to 0; A_0:=A|_{k=0},
\label{eq16} \\
    &B=B_0(1+O(ka)), \quad B_0 u=\varkappa\int_D g_0(x,y)u(y)dy,
\label{eq17} \\
    &B_1u =\varkappa \int_D \frac{\partial g_0(s,y)}{\partial N} u(y)dy
(1+O(k^2a^2)):=\varkappa B_{10}u(1+O(k^2a^2)). \label{eq18}
\end{align}
It follows from  equation \eqref{eq8} that
\begin{equation}
 \label{eq8a}
    u(x)=u_0(x)+\frac{e^{ik|x-x_1|}}{|x-x_1|}\left(\frac{1}{4\pi}
\int_S e^{-ik\beta \cdot t}\sigma(t)dt+\frac{\varkappa}{4\pi}u_1
V_1\right),
\quad |x-x_1|>>a,
\end{equation}
where $V_1$ is the volume of $D=D_1$, $V_1=vol(D_1):=|D_1|$,
$u_1:=u(x_1)$, $\beta:=\frac{x-x_1}{|x-x_1|}$. The point $x_1\in D$
can be chosen as we wish. For one scatterer it is
convenient to choose the origin at the point $x_1$ so that $x_1=0$. \\
We did not keep the factor $e^{-ik\beta \cdot x}$ in the integral
over $D$ because $e^{-ik\beta \cdot x}=1+O(ka)$, and
\begin{equation} \label{eq8b}
     \int_D e^{-ik\beta \cdot y}u(y)dy=u_1V_1(1+O(ka)),
\quad a \to 0.
\end{equation}
However, it will be proved that {\it this factor under the surface
integral can not be dropped} because
\begin{equation} \label{eq8c}
     \int_S e^{-ik\beta \cdot t}\sigma(t)dt=
\int_S \sigma(t)dt-ik\beta_p\int_S t_p\sigma(t)dt+O(a^4),
\end{equation}
where over the repeated indices here and throughout this paper
summation is understood, and the second integral in the right-hand
side of \eqref{eq8c} is $O(a^3)$, as $a \to 0$, that is, it is of
the same order of smallness as the the first integral $Q:=\int_S
\sigma(t)dt$.
The last statement will be proved later. \\
With the notations
\begin{equation} \label{eq8d}
     Q:=\int_S \sigma(t)dt, \quad Q_1:=\int_S e^{-ik\beta \cdot t}
\sigma(t)dt,
\end{equation}
the expression
\begin{equation} \label{eq8e}
     A(\beta,\alpha):=\frac{Q_1}{4\pi}+\frac{\varkappa}{4\pi}u_1 V_1,
\quad V_1:=V:=|D|,\quad u_1:=u(x_1),
\end{equation}
is the scattering amplitude, $\alpha$ is the unit vector in the
direction of the incident wave $u_0 =e^{ik\alpha\cdot x}$, $\beta$
is the unit vector
in the direction of the scattered wave. \\
Let us prove that
\begin{equation} \label{eq8f}
     -ik\beta_p\int_S t_p\sigma(t)dt=O(a^3),
\end{equation}
and therefore, the second integral in the right-hand side of
equation
\eqref{eq8c} cannot be dropped.\\
It follows from  equation \eqref{eq8} that
\begin{equation}
\label{eq19}
     u(x)\sim u_0(x) +g(x,x_1)Q_1+\varkappa g(x,x_1)u(x_1)V_1, \quad
|x-x_1|\geq d\gg a,
\end{equation}
where $\sim$ means asymptotic equivalence as
$a \to 0$.\\
Formula \eqref{eq19} can be used for calculating  $u(x)$ if two
quantities $Q_1$ and $u_1:=u(x_1)$ are found.\\ Let us derive
asymptotic formulas for these quantities as $a\to 0$. Integrate
equation \eqref{eq12} over $S$ and get \begin{equation} \label{eq20}
     Q=2\lambda \int_S u_{0_N}ds + \lambda \int_S A \sigma dt +
2\lambda \int_S B_1 u ds,
\end{equation}
Use formulas \eqref{eq14}-\eqref{eq18}, the following formula (see
\cite{R476}, p.96):
\begin{equation} \label{eq21}
     \int_S A_0 \sigma ds = - \int_S \sigma ds,
\end{equation}
and the Divergence theorem, to rewrite  equation \eqref{eq20} as
\begin{equation} \label{eq22}
     Q=2\lambda\int_D \nabla^2 u_0 dx-\lambda Q +2\lambda\varkappa\int_D dx \nabla^2_x\int_D g(x,y)u(y)dy.
\end{equation}
Since
\begin{equation} \label{eq23}
     \nabla^2 u_0=-k^2 u_0; \quad \nabla^2_x g(x,y)=-k^2g(x,y)-\delta(x-y),
\end{equation}
equation \eqref{eq22} takes the form
\begin{equation} \label{eq24}
     (1+\lambda)Q=2\lambda \nabla^2 u_0(x_1)V_1 -2\lambda k^2\varkappa
\int_D dx \int_D g(x,y) u dy - 2\lambda \varkappa\int_D u(x) dx.
\end{equation}
Let us use the following estimates:
\begin{align}
     &\int_D u(x)dx=u_1V_1 (1+o(1)), \quad a \to 0; \quad
u_1:=u(x_1), \label{eq25} \\
     &\int_D dx\int_D g(x,y) u(y)dy = \int_D dy u(y)\int_Ddx g(x,y)=
O(a^5), \label{eq26} \\
     &\int_D g(x,y)dx = O(a^2), \quad \forall y \in D. \label{eq27}
\end{align}
From  equations \eqref{eq24}-\eqref{eq27} it follows that
\begin{equation}\label{eq28}
Q \sim \frac{2\lambda}{1+\lambda} V_1 \nabla^2 u_{01} -
\frac{2 \lambda \varkappa}{1+\lambda}V_1 u_1, \quad a \to 0,
\end{equation}
where
\begin{equation}\label{eq29}
\nabla^2 u_{01}=\nabla^2u_0(x)|_{x=x_1}.
\end{equation}
Let us now integrate equation \eqref{eq8} over $D$ and use estimate
\eqref{eq25} to obtain
\begin{equation} \label{eq30}
     u_1 V_1=u_{01}V_1 +\int_S dt \sigma(t)\int_D g(x,t)dx+
\varkappa\int_D dy u(y)\int_D g(x,y) dx.
\end{equation}
If $D$ is a ball of radius $a$, then one can easily check that
\begin{equation} \label{eq31}
     \int_D g(x,t)dx \sim \int_D g_0(x,t)dx=\frac{a^2}{3}, \quad |t|=a,
\quad a \to 0.
\end{equation}
In general, one has
\begin{equation} \label{eq32}
     \int_D g(x,y)dx = O(a^2), \quad y \in D,\quad a \to 0.
\end{equation}
If $D$ is a ball of radius $a$, then equations
\eqref{eq30}-\eqref{eq32} imply
\begin{equation} \label{eq33}
     u_1=u_{01}+Q\frac{a^2}{3\frac{4\pi a^3}{3}}+\varkappa u_1 O(a^2),
\quad a \to 0.
\end{equation}
Consequently,
\begin{equation} \label{eq34}
     u_1 \sim u_{01}+O(a^2), \quad a \to 0,
\end{equation}
because $Q=O(a^3)$.

Indeed, from  equations \eqref{eq28} and \eqref{eq34} one gets
\begin{equation} \label{eq35}
     Q \sim V_1(1-\rho) [\nabla^2u_{01}-\varkappa u_{01}],
\end{equation}
where we took into account that
\begin{equation} \label{eq36}
     \frac{2\lambda}{1+\lambda}=1-\rho,
\end{equation}
the relation $u_1\sim u_{01}$ as $a\to 0$, see  equation
\eqref{eq34}, and neglected the terms of higher order of smallness.
It follows from  equation \eqref{eq35} that
\begin{equation} \label{eq37}
     Q=O(a^3).
\end{equation}
From  equations \eqref{eq34} and \eqref{eq35} one obtains
\begin{equation} \label{eq38}
     u_1 \sim u_{01}, \quad a \to 0.
\end{equation}
Let us now estimate $Q_1$. One has
\begin{equation} \label{eq39}
     Q_1=\int_S \sigma(t)dt-ik\beta_p\int_S t_p\sigma(t)dt,
\end{equation}
up to the terms of the higher order of smallness as $a\to 0$, and
summation is understood over the repeated indices. It turns out that
the integral
\begin{equation} \label{eq39a}
     I:=\int_S t_p \sigma(t)dt
\end{equation}
is of the same order, namely  $O(a^3)$, as $Q=\int_S \sigma(t)dt$.\\
Let us check that the integral
$$J:=\int_S dt t_p\frac{\partial}{\partial
N}\int_D g(t,y)u(y)dy=O(a^4)$$
as $a\to 0$, and, therefore, can be neglected compared with $I$.
Indeed, $u=O(1)$, $\int_D\frac{\partial}{\partial N}g(t,y)dy=O(a)$,
and $\int_St_pdt=O(a^3)$. Thus, $J=O(a^4)$.

Define the function $\sigma_q$, $q=1,2,3,$ as the unique solution to
the equation
\begin{equation} \label{eq47}
\sigma_q=\lambda A\sigma_q-2\lambda N_q.
\end{equation}
Since $\lambda=(1-\rho)/(1+\rho)$, and $\rho>0$, one concludes that
$\lambda\in (-1,1)$, and it is known (see, for example, \cite{R476})
that the operator $A$ is compact in $L^2(S)$ and does not have
characteristic values in the interval $(-1,1)$. This and the
Fredholm alternative imply that equation \eqref{eq47} has a solution
and this solution is unique.

Let us prove that $\int_S\sigma_q(t)dt=O(a^3).$ To do this,
integrate equation \eqref{eq47} over $S$, take into account formula
\eqref{eq21}, the relation $(A-A_0)\sigma_q=O(a^3)$, and obtain
$$(1+\lambda)\int_S\sigma_q(t)dt=-2\lambda \int_SN_qdt +O(a^3)=O(a^3),$$
because $\int_SN_qdt=0$ by the Divergence theorem.

Define the tensor
\begin{equation} \label{eq48}
\beta_{pq}:=\beta_{pq}(\lambda):={V_1}^{-1}\int_St_p\sigma_q(t)dt,
\qquad
p,q=1,2,3.
\end{equation}
This tensor is similar to the tensor $\beta_{pq}$ defined in
\cite{R476}, p. 62, by a similar formula with $\lambda=1$. In this
case $\beta_{pq}$ is the magnetic polarizability tensor of a
superconductor $D$ placed in a homogeneous magnetic field directed
along the unit Cartesian coordinate vector $e_q$ (see \cite{R476},
p. 62). In \cite{R476} analytic formulas are given for calculating
$\beta_{pq}$ with an arbitrary accuracy.

One may neglect the term $B_1u$ in equation \eqref{eq12} (because
this term is $O(a^4)$), take into account definition \eqref{eq48},
and get
\begin{equation} \label{eq49}
\int_St_p\sigma(t)dt= -\beta_{pq}\frac{\partial u_0}{\partial x_q} V,
\end{equation}
where $V:=V_1$, and summation is understood over $q$.

Consequently, one can rewrite \eqref{eq39} as
\begin{equation} \label{eq50}
   Q_1=(1-\rho)V_1[\nabla^2(u_0(x_1)-\varkappa
u_0(x_1)]+ik\beta_{pq}\frac{\partial u_0}{\partial
x_q}\beta_p V_1, \quad \beta:=\frac{x-x_1}{|x-x_1|},
\end{equation}
and $(x)_p:=x\cdot e_p$ is the $p-$th Cartesian coordinate of the
vector $x$.

Formula \eqref{eq8a} can be written as
\begin{equation}\label{eq51}
        u(x)=u_0(x)+g(x,x_1)\Big(
(1-\rho)[\nabla^2u_0(x_1)-\varkappa u_0(x_1)]+\\
 ik\beta_{pq}\frac{\partial
u_0(x_1)}{\partial x_{1,q}}\beta_p +\varkappa u_0(x_1) \Big)V_1.
\end{equation}
Here one sums  over the repeated indices, $|x-x_1|>>a$, and
$\frac{\partial u_0(x_1)}{\partial x_{1,q}}:=\frac{\partial u_0(y)}
{\partial y_q}|_{y=x_1}$, $q=1,2,3$, $y=(y_1, y_2, y_3)$.

Formulas \eqref{eq35},\eqref{eq37},\eqref{eq38} are valid for small
$D$ of arbitrary shape. Let us formulate the results of this Section
in the following theorem.

{\bf Theorem 6.1.} {\it  Assume that $ka\ll 1$, $k_1, k,$ and $\rho$
are positive constants. Then the scattering problem
\eqref{eq1}-\eqref{eq5} has a unique solution. This solution has the
form \eqref{eq8} and can be calculated by formula \eqref{eq51} in
the region $|x-x_1|>>a$ up to the terms of order $O(a^4)$ as $a\to
0$, where $a=0.5 diam D$, $\varkappa= k_1^2-k^2$, $V_1=vol D$,
$\beta=\frac{x-x_1}{|x-x_1|}$,  $\beta_{pq}$ is defined in equation
\eqref{eq48}, and $O(a^4)$ does not depend on $x$.}

\section{Wave scattering by many small bodies}
Assume  that
the distribution of small bodies is given by equation \eqref{eq6},
and that there are $M=M(a)$ non-intersecting small bodies $D_m$ of
size $a$. For simplicity we assume that $D_m$ is a ball of radius a,
centered at $x_m$. There is an essential novel feature in the
theory, developed in this paper compared with the one developed in
\cite{R509},\cite{R536}, \cite{R595}, namely,  the scattered field
was much larger, as $a \to 0$ in the above papers. For example, for
the impedance boundary condition, $u_N=\zeta u$ on $S$, the
scattered field is $O(a^2)$, and for the Dirichlet boundary
condition, $u=0$ on $S$, the scattered field is $O(a)$.

For the Neumann boundary condition the scattered field is $O(a^3)$.
We have the same order of smallness of the scattered field,
$O(a^3)$, for the problem with the transmission boundary condition
because $V_1=O(a^3)$. The basic role in  Section 3 is played by
formula \eqref{eq51}. We assume that the distance $d$ between
neighboring bodies (particles) is  much larger than $a$, $d>>a$, but
there can be many small particles on the wavelength, and the
interaction of the scattered waves (multiple scattering) is
essential and cannot be neglected.

This assumption effectively means that the function $N(x)$ in
\eqref{eq6} has to be small, $N(x)<<1$. Indeed, if on a segment of
unit length there are small particles placed at a distance $d$
between neighboring particles, then there are $O(\frac{1}{d})$
particles on this unit segment, and $O(\frac{1}{d^3})$ in a unit
cube $C_1$. Since $V=O(a^3)$, by formula \eqref{eq6} one gets
$$  \frac{1}{O(a^3)}\int_{C_1} N(x)dx = O(\frac{1}{d^3}).$$
Therefore $d>>a$ can hold only if
$(\int_{C_1}N(x)dx)^{\frac{1}{3}}=O(\frac{a}{d})<<1$.

Let us look for the (unique) solution to problem
\eqref{eq1}-\eqref{eq5} with $1\leq m\leq M=M(a)$ of the form
\begin{equation} \label{eq41a}
    u(x)=u_0(x)+\sum_{m=1}^M \int_{S_m} g(x,t)\sigma_m(t)dt+
\sum_{m=1}^M \varkappa_m\int_{D_m}g(x,y)u(y)dy.
\end{equation}
Keeping the main terms in this equation, as $a\to 0$, one gets
\begin{align} \label{eq42}
    &u(x)=u_0(x)+\sum_{m=1}^M g(x,x_m)\left(Q_m-ik\frac{(x-x_m)_p}
{|x-x_m|}\int_{S_m}t_p\sigma_m(t)dt\right)+ \nonumber\\
    &\qquad +\sum_{m=1}^M \varkappa_m g(x,x_m)u_e(x_m)V_m,
    \qquad Q_m:=\int_{S_m}\sigma_m(t)dt, \quad a \to 0,
\end{align}
where we have
used formula \eqref{eq51} for the scattered field by every small
particle, replaced $u_0$ by the effective field $u_e$, acting on
every particle, and took into account that
$\beta:=\beta_m:=\frac{x-x_m}{|x-x_m|}$. By $(x-x_m)_p$ the $p$-th
component of vector $(x-x_m)$ is denoted.

The effective (self-consisted) field $u_e$, acting on $j$-th
particle, is defined as:
\begin{align} \label{eq43}
    &u_e(x)=u_0(x)+\sum_{m=1,m\neq j}^M g(x,x_m)\left(
(1-\rho_m)[\nabla^2u_e(x_m)-\varkappa_m u_e(x_m)]+\right.\nonumber\\
&\left.ik\beta^{(m)}_{pq}\frac{\partial
u_e}{\partial x_q}\frac{(x-x_m)_p}{|x-x_m|} \right)V_m
 +\sum_{m=1,m\neq j}^M \varkappa_m
g(x,x_m)u_e(x_m) V_m, \quad |x-x_j| \sim a.
\end{align}
Setting $x=x_j$ in equation \eqref{eq43}, one gets {\it a linear
algebraic system for the unknowns $u_j:=u_e(x_j), 1\leq j\leq M$,
and $\frac{\partial u_e(x_j)}{\partial x_{j,p}}$.} Here $x_{j,p}$ is
the $p-$th component of the vector $x_j$, $p=1,2,3$. Differentiating
\eqref{eq43} with respect to $x_{j,p}$, $p=1,2,3,$ and then setting
$x=x_j$, one obtains {\it a linear algebraic system for the $4M$
unknowns $u_j$ and $\frac{\partial u_e(x_j)}{\partial x_{j,p}}$,
$1\leq j\leq M$, $1\leq p\leq 3$.}

This linear algebraic system one gets if one solves by a collocation
method   the following integral equation
\begin{align}\label{eq46}
    &u(x)=u_0(x)+\int_D
g(x,y)\Big[(1-\rho)(\nabla^2-K^2(y)+k^2)u(y)+\nonumber\\
    &ik\beta_{pq}(y,\lambda)\frac{\partial u(y)}{\partial
y_q}\frac{(x-y)_p}{|x-y|}+(K^2(y)-k^2)u(y)\Big]N(y)dy.
\end{align}
In the above equation the function $\beta_{pq}(y,\lambda)$ is
defined as
$$\beta_{pq}(y,\lambda)=\lim_{a\to 0} \frac{\sum_{x_m\in
\Delta_p}\beta^{(m)}_{pq}}{\mathcal{N}(\Delta_p)},$$
where  $y=y_p\in \Delta_p$, and tensor
$\beta_{pq}^{(m)}=\beta_{pq}^{(m)}(\lambda)$ is defined in
\eqref{eq48}. Convergence of the collocation method was proved in
\cite{R595}.

 {\it Equation \eqref{eq46} is  a non-local integrodifferential
equation for the limiting effective field in the medium in which
many small bodies are embedded}.

This is a novel result. The original scattering problem
\eqref{eq1}-\eqref{eq5} has been formulated in terms of local
differential operators.

In the derivation of equation  \eqref{eq46} from equation
\eqref{eq43} we have assumed that $\rho_m=\rho$ does not depend on
$m$, took into account that $\varkappa_m^2$ becomes in the limit
$K^2(y)-k^2$, and denoted by $K^2(y)$ a continuous function in $D$
such that $K^2(x_m)=k^2_m$. As $a \to 0$ the function $K^2(y)$ is
uniquely defined because the set $\{x_m\}_{m=1}^{M(a)}$ becomes
dense in $D$ as $a \to 0$.

To derive equation \eqref{eq46} from equation  \eqref{eq43} we argue
as follows. Consider a partition of $D$ into a union centered at the
points $y_p$ of $P$ non-intersecting cubes $\Delta_p$, of size
$b(a)$, $b(a)>>d$, so that each cube contains many small bodies,
$lim_{a \to 0}b(a)=0$. Let us demonstrate the passage to the limit
$a\to 0$ in the sums in equation \eqref{eq43} using the first sum as
an example. Write the first sum in \eqref{eq43} as
\begin{align} \label{eq52}
    &\sum_{m\neq j}g(x,x_m)(1-\rho_m)[\nabla^2u_e(x_m)-
\kappa_m u_e(x_m)]V_m \nonumber \\
    &=\sum_{p=1}^P g(x,y_p)(1-\rho_p)[\nabla^2u_e(y_p)-\kappa_p
u_e(y_p)]V_m \sum_{x_m \in \Delta_p} 1 \nonumber \\
    &=\sum_{p=1}^P
g(x,y_p)(1-\rho_p)[\nabla^2u_e(y_p)-\kappa_pu_e(y_p)]N(y_p)|\Delta_p|
(1+o(1)),
\end{align}
where we have used formula \eqref{eq6}, took into account that
$diam \Delta_p \to 0$ as $a\to 0$, wrote formula  \eqref{eq6} as
\begin{equation} \label{eq53}
    V\sum_{x_m \in \Delta_p}1=V\mathcal{N}(\Delta_p)=
N(y_p)|\Delta_p|(1+o(1)), \quad a\to 0,
\end{equation}
and used the Riemann integrability of the functions involved, which
holds, for example, if these functions are continuous. By $\rho_p$
we denote the value $\rho(y_p)$, where $\rho(y)$ is a continuous
function.

The sum in \eqref{eq52} is the Riemannian sum for the integral
\begin{equation}\label{eq54}
\int_Dg(x,y) (1-\rho (y))[\nabla^2u(y)-K^2(y)u(y)+k^2 u(y)]N(y)dy.
\end{equation}
Similarly one treats the other sums in \eqref{eq43} and obtains in
the limit $a\to 0$ equation \eqref{eq46}.

Let us formulate the results of this Section in the following theorem.

{\bf Theorem 7.1.} {\it Assume that conditions \eqref{eq6} and
\eqref{eq7} hold. Then, as $a\to 0$, the effective field, defined by
equation \eqref{eq43}, has a limit $u(x)$. The function  $u(x)$
solves equation \eqref{eq46}.}
$$ $$
\section{Conclusions}
In this paper analytic formulas for the scattering amplitudes
for wave scattering by a single small particle are derived
for various boundary conditions: the Dirichlet, Neumann, impedance,
and transmission ones.

The equation for the effective field in the medium, in which many small
particles are embedded, is derived in the limit $a\to 0$. The
physical assumptions are such that the multiple scattering effects are
not negigible, but essential. The derivations are rigorous.

On the basis of the developed theory efficient numerical
 methods are proposed for solving many-body wave scattering problems in
the case of small scatterers.

\end{document}